\documentstyle[12pt]{article}
\begin{document}
\begin{center}{\large\bf Dark Energy and  Dilaton Cosmology }\\
 \vskip 0.15 in
H.Q.Lu{\footnote{$Alberthq_-Lu@hotmail.~com$}} Z.G.Huang W.Fang  K.F.Zhang\\

{\it Department of Physics, \\Shanghai University, Shanghai
200436, China}\\
\vskip 0.3 in
\begin{minipage}{5.5in}
{\small ~~We studied the dilaton cosmology based on Weyl-Scaled
induced gravity. The potential of dilaton field is taken as
exponential form. An analytical solution of Einstein equation is
found. The dilaton can be a candidate for dark energy that can
explain the accelerated universe. The structure formation is also
considered. We find the the evolutive equation of density
perturbation, and its growth is quicker than the one in standard
model which is consistent with the constraint from CMBR
measurements
\\{\bf PACS:}98.80.Cq,04.50.+h}
\end{minipage}
\end{center}
\vskip 0.3 in \begin{center}\textbf{I.Introduction}\end{center}
~~~~The observation evidence for accelerating universe has been
one of the central themes of modern cosmology for the past few
years. The explanation of these observations in the framework of
standard cosmology requires an exotic form of energy which
violates the strong energy condition. A variational of scalar
field model has been conjectured for this purpose including
quintessence[1], K-essence[1] and recently tachyonic scalar fields
[1].
\par Among these models, the important one is tachyon model.
The role of tachyon field in string theory in cosmology has been
widely studied. However, the model of inflation with a single
tachyon field generates larger anisotropy and has troubles in
describing the forming of the universe[2]. A successful extended
inflation(shortened EI) can occur in Weyl-Scaled induced
gravity(also was called induced gravity in the Einstein frame).
During the EI epoch, the universe expands according to a power
law. It has been shown that this power-law expansion is fast
enough to resolve cosmological puzzles, but slow enough to
percolate the false vacuum via nucleation of true vacuum
bubbles[3]. The EI models also predict the primordial spectral
indices of density perturbations, and the values of spectral
indices is close to the observations[4]. The models predict the
fluctuation as follows
\begin{equation}(\frac{\delta\rho}{\rho})\sim10^{-5}\end{equation}
 one can find the fluctuation $(\frac{\delta\rho}{\rho})$ well within
currently observed limits[3]. A successful EI model leads to a
result that we regard the dilaton field of included gravity as
theoretical models of dark energy.
 \begin{center}\textbf{II.Cosmological Dynamics In The Presence of Dilaton
 Field}\end{center}
 ~~~~Let us consider the action of Jordan-Brans-Dicke theory
 \begin{equation}S=\int{d^4x\sqrt{-\gamma}[\phi
 R+\omega\gamma^{\mu\nu}\frac{\partial_\mu\phi\partial_\nu\phi}{\phi}-\Lambda(\phi)+L_{fluid}(\psi)}]\end{equation}
where the lagrangian density of cosmic fluid
$L_{fluid}(\psi)=\frac{1}{2}\gamma^{\mu\nu}\partial_\mu\psi\partial_\nu\psi-V(\psi)$,
$\gamma$ is the determinant of the Jordan metric tensor
$\gamma_{\mu\nu}$, $\omega$ is the dimensionless coupling
parameter, $R$ is the contracted $R_{\mu\nu}$. The metric sign
convention is(-,+,+,+). The quantity $\Lambda(\phi)$ is a
nontrivial potential of $\phi$ field. When $\Lambda(\phi)\neq0$
the action of Eq.(2) describes the induced gravity. The energy
density of cosmic fluid
$\widetilde{\rho}=\frac{1}{2}(\frac{d\psi}{d\widetilde{t}})^2+V(\psi)$
and the pressure $
\widetilde{p}=\frac{1}{2}(\frac{d\psi}{d\widetilde{t}})^2-V(\psi).$
\par However it is often useful to write the action in terms of the conformally related Einstein metric. We introduce the dilaton field $\sigma$ and
conformal transformation as follows
\begin{equation}\phi=\frac{1}{2}e^{\alpha\sigma}\end{equation}
\begin{equation}\gamma_{\mu\nu}=e^{-\alpha\sigma}g_{\mu\nu}\end{equation}
where $\alpha^2=\frac{\kappa^2}{2\omega+3}$. $\kappa^2=8\pi G$ is
taken to be one for convenience.
\\~~~~~The action(2) becomes Eq.(5) by performing the conformal
transformation Eq.(3) and Eq.(4)
\begin{equation}S=\int{d^4x\sqrt{-\gamma}[\frac{1}{2}\tilde{R}(g_{\mu\nu})+\frac{1}{2}g^{\mu\nu}\partial_\mu\sigma\partial_\nu\sigma-W(\sigma)+\tilde{L}_{fluid}(\psi)}]\end{equation}
where
$\tilde{L}_{fluid}(\psi)=\frac{1}{2}g^{\mu\nu}e^{-\alpha\sigma}\partial_\mu\psi\partial_\nu\psi-e^{-2\alpha\sigma}V(\psi)$
\par The transformation Eq.(3) and Eq.(4) are well defined for some
$\omega$ as $-\frac{3}{2}<\omega<\infty$. The conventional
Einstein gravity limit occurs as $\sigma\rightarrow 0$ for an
arbitrary $\omega$ or $\omega\rightarrow\infty$ with an arbitrary
$\sigma$.
\par The nontrivial potential of the $\sigma$ field,
$W(\sigma)$ can be a metric scale form of $\Lambda(\phi)$.
Otherwise, one can start from Eq.(5), and define $W(\sigma)$ as an
arbitrary nontrivial potential. $g_{\mu\nu}$ is the pauli metric.
Cho and Damour et.al pointed out  that  the pauli metric can
represent the massless spin-two graviton in induced gravitational
theory[5]. Cho also pointed out that in the compactification of
Kaluza-Klein theory, the physical metric must be identified as the
pauli metric because of the the wrong sign of the kinetic energy
term of the scalar field in the Jordan frame. The dilaton field
appears in string theory naturally.
\par By varying the action Eq.(5), one can obtain the field equations
of Weyl-scaled induced gravitational theory.
$$R_{\mu\nu}-\frac{1}{2}g_{\mu\nu}R=-\frac{1}{3}\{[\partial_\mu\sigma\partial_\nu\sigma-\frac{1}{2}g_{\mu\nu}\partial_\rho\sigma\partial^\rho\sigma]
-g_{\mu\nu}W(\sigma)$$
\begin{equation}+e^{-\alpha\sigma}[\partial_\mu\psi\partial_\nu\psi-\frac{1}{2}g_{\mu\nu}\partial_\rho\psi\partial^\rho\psi]
-g_{\mu\nu}e^{-2\alpha\sigma}V(\psi)\}\end{equation}
\begin{equation}\Box\sigma=\frac{dW(\sigma)}{d\sigma}-\frac{\alpha}{2}e^{-2\alpha\sigma}g^{\mu\nu}\partial_\mu\psi\partial_\nu\psi
-2\alpha e^{-2\alpha\sigma}V(\psi)\end{equation}
\begin{equation}\Box\psi=-\alpha g_{\mu\nu}\partial_\mu\psi\partial_\nu\sigma+e^{-\alpha\sigma}\frac{dV(\psi)}{d\psi}\end{equation}
The energy-momentum tensor $T_{\mu\nu}$ of cosmic fluid is
\begin{equation}T_{\mu\nu}=(\rho+p)U_\mu U_\nu+pg_{\mu\nu}\end{equation}
where the density of energy
\begin{equation}\rho=\frac{1}{2}\dot{\psi}^2+e^{-\alpha\sigma}V(\psi)\end{equation}
the pressure
\begin{equation}p=\frac{1}{2}\dot{\psi}^2-e^{-\alpha\sigma}V(\psi)\end{equation}
$\rho$ and $p$ are related to their directly measurable
counterparts by $\rho=e^{-\alpha\sigma}\widetilde{\rho},
p=e^{-\alpha\sigma}\widetilde{p}.$
 We work in  R-W metric
\begin{equation}ds^2=-dt^2+a^2(t)(dx^2+dy^2+dz^2)\end{equation}
and we consider that  $\rho,p$  and  $\sigma$  depend only on
time. So, according to Eqs.(6)-(12), we can obtain
\begin{equation}(\frac{\dot{a}}{a})^2=\frac{1}{3}[\frac{1}{2}\dot{\sigma}^2+W(\sigma)+e^{-\alpha\sigma}\rho]\end{equation}
\begin{equation}\ddot{\sigma}+3H\dot{\sigma}+\frac{dW}{d\sigma}=\frac{1}{2}\alpha e^{-\alpha\sigma}(\rho-3p)\end{equation}
\begin{equation}\dot{\rho}+3H(\rho+p)=\frac{1}{2}\alpha\dot{\sigma}(\rho+3p)\end{equation}
\par Exponential  potential attract much attention because they can
be derived from the effective interaction in string theory and
Kaluza-Klein theory. Their roles in cosmology have also been
widely investigated[6].
\par In this paper we assume that
$W(\sigma)=Ae^{-\beta\sigma}$. From Eq.(14) we can obtain the
solution in dilaton energy-dominanted epoch,
\begin{equation}\sigma=\frac{2}{\beta}ln[\sqrt{\frac{A}{2(\frac{6}{\beta^2}-1)}}|\beta|t]\end{equation}
\par For radiation $\rho_r=3p_r$, we get
$\rho_r\propto\frac{e^{\alpha\sigma}}{a^4}$ from Eq.(15). For
matter with $p_m=0$, we get
$\rho_m\propto\frac{e^{\frac{1}{2}\alpha\sigma}}{a^3}$ from
Eq.(15). The effective energy density of dilaton field is
$\rho_\sigma=\frac{1}{2}\dot{\sigma}^2+W(\sigma)$, and the
effective pressure is
$p_\sigma=\frac{1}{2}\dot{\sigma}-W(\sigma)$.
\par At the very
large cosmological scale , the contributions from matter and
radiation in Eq.(13) become negligible compared with the dilaton
field. In order to see this more clearly, we take the example that
when the equation of state of the dark energy(dilaton energy) is
constant and $\frac{p_\sigma}{\rho_\sigma}$ must be less than
$-\frac{1}{3}$, the expansion of the universe accelerates. Thus
the dark energy component will evolve with $\rho_\sigma\propto
a^{-3(1+\frac{p_\sigma}{\rho_\sigma})}$, which dissipates slower
than radiation($e^{-\alpha\sigma}\rho_r\propto a^{-4} $) and
matter($e^{-\alpha\sigma}\rho_m\propto
\frac{t^{-\frac{\alpha}{\beta}}}{a^3} $).
$\alpha^2=\frac{1}{2\omega+3}<10^{-3}$ which is constrained by
present-day solar system test[4]. So at late time the dilaton
energy ultimately dominates in universe, and the Eq.(13) becomes
\begin{equation}(\frac{\dot{a}}{a})^2=\frac{1}{3}[\frac{1}{2}\dot{\sigma}^2+W(\sigma)]\end{equation}
From Eqs.(14) and (17), we obtain
\begin{equation}a(t)=a_0(\frac{t}{t_0})^{\frac{2}{\beta^2}}\end{equation}
and solution of $\sigma$ is Eq.(16). After investigating stability
of the solutions (16) and (18), we can find that the Eqs.(16) and
(18) are stable for $\beta^2<2$. This is similar to $\Phi$CDM[8].
\\The effective energy density of dilaton is
\begin{equation}\rho_\sigma=\frac{Ae^{-\beta\sigma}}{1-\frac{\beta^2}{6}}\end{equation}
The effective pressure is
\begin{equation}p_\sigma=\frac{A(\frac{\beta^2}{3}-1)e^{-\beta\sigma}}{1-\frac{\beta^2}{6}}\end{equation}
\\From Eqs.(18) and (19), one can get
\begin{equation}\frac{p_\sigma}{\rho_\sigma}=(\frac{\beta^2}{3}-1)\end{equation}
\par When $\beta^2$ is smaller than 2, Eq.(21) shows that
$\rho_\sigma+3p_\sigma<0$ and the universe is undergoing a phase
of accelerated expansion. It is clear that the
$\rho_\sigma+p_\sigma$ is greater than zero from Eq.(21). When
$\beta$ approximates zero, $\rho_\sigma=-p_\sigma$.
\begin{center}\textbf{III.The Discussion of Cosmological Structure Formation In Dilaton Cosmology}\end{center}

\par We will find that the growth of density perturbation is quicker than the one in standard model.   Eq.(13) becomes
\begin{equation}(\frac{\dot{a}}{a})^2=\frac{1}{3}[\frac{1}{2}\dot{\sigma}^2+W(\sigma)+ e^{-\alpha\sigma}\rho]\end{equation}
The cosmic fluid can be regard as the nonrelativistic matter $p\ll
\rho$. From Eqs.(22),(14)and (15), we obtain
\begin{equation}\sigma=ln\sigma_0+\frac{2}{\beta}ln[\sqrt{\frac{A}{2(\frac{6}{\beta^2}-1)}}|\beta|t]\end{equation}
\begin{equation}a=a_e(\frac{t}{t_e})^{\frac{2}{3}-\frac{\alpha}{3\beta}}\end{equation}
\begin{equation}\rho=\rho_e(\frac{t}{t_e})^{-2+\frac{2\alpha}{\beta}}\end{equation}
The radiation density equals matter density at $t_e$,
correspondingly the cosmological scalar is $a_e$ and the matter
density is $\rho_e$. In the early time of the matter
energy-dominated epoch,
$ln\sigma_0>>\frac{2}{\beta}ln[\sqrt{\frac{A}{2(\frac{6}{\beta^2}-1)}}|\beta|t]$
and thus $W(\sigma)$ approximates to a constant
$A{\sigma_0}^{-\beta}$. However the terms of
$e^{-\alpha\sigma}\rho$ decreases as $t^{-2}$, so the Universe
becomes dominated by the dilaton energy little by little.
\\For $p=0$ the
Eq.(15) becomes
\begin{equation}\dot{\rho}+3H\rho=\frac{1}{2}\alpha\dot{\sigma}\rho\end{equation}
The Eq.(26) of motion of fluid can also be written as follows
\begin{equation}\frac{\partial[e^{-\alpha\sigma}\rho(t)]}{\partial t}+\nabla\cdot[e^{-\alpha\sigma}\rho \vec{V}]
=-\frac{1}{2}\alpha\dot{\sigma}\rho
e^{-\alpha\sigma}\end{equation}
\\In standard model, the equation of motion of comic fluid is
$$\dot{\rho}+3H\rho=0$$
namely \begin{equation}\frac{\partial\rho}{\partial
t}+\nabla\cdot(\rho\vec{V})=0 \end{equation}The Eq.(27) is
different from Eq.(28) due to the effect of dilaton field. when
$\dot{\sigma}=0$, Eq.(27) becomes Eq.(28)
\par Considering a fluctuating region which is by far smaller
than the universe, we can deal with it by the following theory of
fluctuation[7]. The Euler equation is
\begin{equation}\frac{\partial\vec{V}}{\partial t}+(\vec{V}\cdot\nabla)\vec{V}=-\frac{1}{\rho}\nabla p+\vec{g}\end{equation}
where $\vec{V}$ is the speed of fluid, $\vec{g}$ is the intensity
gravitational field exerted by fluid, and $\vec{g}$ satisfies the
following equations
\begin{equation}\nabla\times\vec{g}=0\end{equation}
\begin{equation}\nabla \vec{g}=-4\pi G\rho\end{equation}
According to  Eqs.(22)(27)(29)(30)(31), we can obtain a simple
spacial homogeneous solution

\begin{equation}a(t)=a_e(\frac{t}{t_e})^{\frac{2}{3}-\frac{\alpha}{3\beta}}\end{equation}
\begin{equation}\rho(t)=\rho_e(\frac{t}{t_e})^{-2+\frac{2\alpha}{\beta}}\end{equation}
\begin{equation}\vec{V}=\vec{r}~[\frac{\dot{a}(t)}{a(t)}]\end{equation}
\begin{equation}\vec{g}=-\vec{r}~(\frac{4\pi G\rho}{3})\end{equation}
Eq.(34) is a solution that satisfies the Hubble law. Now we solve
the perturbation solution. One can add the zero series solution
$\rho,p,\vec{V},\vec{g}$ to the perturbation solution
$\rho_1,p_1,\vec{V_1},\vec{g_1}$. According to Eqs.(27)and(29),
one can get the first series approximate expression
\begin{equation}\frac{\partial\rho_1}{\partial t}+\frac{3\dot{a}}{a}\rho+\frac{\dot{a}}{a}(\vec{r}\cdot\nabla)
\rho_1+\rho\nabla\cdot
\vec{V_1}-\frac{1}{2}\alpha\dot{\sigma}\rho_1=0\end{equation}
\begin{equation}\frac{\partial \vec{V_1}}{\partial t}+\frac{\dot{a}}{a}\vec{V_1}+\frac{\dot{a}}{a}(\vec{r}\cdot\nabla)\vec{V_1}
=-\frac{1}{\rho}\nabla p+\vec{g_1}\end{equation}
\\From Eq.(30) and Eq.(31), one can obtain respectively
\begin{equation}\nabla\times \vec{g_1}=0\end{equation}
\begin{equation}\nabla\cdot \vec{g_1}=-4\pi G\rho_1\end{equation}
Assuming the perturbation is adiabatic, one knows that $p_1$ is
decided by the following equation
\begin{equation}p_1=V_s^2\rho_1\end{equation}
where $V_s$ is sonic speed. Because Eqs.(36)-(39) is spacial
homogeneous, one expects a plane wave solution. In fact, according
to the spacial dependence
\begin{equation}\rho_1(\vec{r},t)=\rho_1(t)exp\{\frac{i\vec{r}\cdot \vec{q}}{a(t)}\}\end{equation}
and the analogous expression of $\vec{V_1}$  and $\vec{g_1}$, one
can obtain these solutions(when the factor $\frac{1}{a}$ appears
in the exponential, it means that the wavelength is enlarged by
the expansion of universe). Substituting Eq.(41), the analogous
expression of $\vec{V_1}$ and $\vec{g_1}$ into Eqs.(36)-(39), one
can obtain
\begin{equation}\frac{\partial\rho_1}{\partial t}+\frac{3\dot{a}}{a}\rho_1+ia^{-1}\vec{q}\cdot\vec{V_1}\rho-\frac{1}{2}\alpha\dot{\sigma}\rho_1=0\end{equation}
\begin{equation}\dot{\vec{V_1}}+\frac{\dot{a}}{a}\vec{V_1}=-\frac{iV^2_s}{\rho a}q\rho_1+\vec{g_1}\end{equation}
\begin{equation}\vec{q}\times\vec{g_1}=0\end{equation}
\begin{equation}i\vec{q}\cdot\vec{g_1}=-4\pi G\rho_1a\end{equation}
Eqs.(44) and (45) have the obvious solution
\begin{equation}\vec{g_1}=\frac{4\pi G\rho_1a\vec{q}\vec{i}}{q^2}\end{equation}
For solving the equation of motion, it is convenient to separate
the $\vec{V}$ into the parallel part and vertical part of
$\vec{q}$
\begin{equation}\vec{V_1}(t)=\vec{V_{1\perp}}(t)+i\vec{q}\varepsilon(t)\end{equation}

where $\vec{q}\cdot\vec{V_{1\perp}}=0$
\begin{equation}\varepsilon=\frac{-i\vec{q}\cdot\vec{V_1}}{q^2}\end{equation}
Introducing the relative change of density
\begin{equation}\delta(t)=\frac{\rho_1(t)}{\rho(t)}\end{equation}
Eq.(43) is decomposed into two separate equations
\begin{equation}\dot{\vec{V_{1\perp}}}+\frac{\dot{a}}{a}\vec{V_{1\perp}}=0\end{equation}
\begin{equation}\dot{\varepsilon}+\frac{\dot{a}}{a}\varepsilon=(-\frac{V_s^2}{a}+\frac{4\pi G\rho a}{q^2})\delta\end{equation}
Taking Eq.(49) into Eq.(42), one can obtain
\begin{equation}\varepsilon=\frac{\dot{\delta}a}{q^2}\end{equation}
Taking $\varepsilon$ and $\dot{\varepsilon}$ obtained from Eq.(52)
into Eq.(51), one can obtain
\begin{equation}\ddot{\delta}+\frac{2\dot{a}}{a}\dot{\delta}+(\frac{V_s^2q^2}{a^2}-4\pi G\rho)\delta=0\end{equation}
It was a basic equation which controlled the increase or decline
of gravitational agglomeration in  universe. When the energy
density of radiation decreased below the rest mass density, the
matter epoch began, and the above nonrelativistic perturbation
theory became valid.
\par We consider the solution when $p=0$. It is well known that the
galaxies were exerted by the perturbation of matter density. How
many times have the original perturbation increased from
recombination time to present? For answering the question, we
ignore $\frac{V_s^2q^2}{a^2}$ in order to simplify the above
equation, so, Eq.(53) becomes
\begin{equation}\ddot{\delta}+\frac{2\dot{a}}{a}\dot{\delta}-4\pi G\rho\delta=0\end{equation}
Substitute  Eq.(32)(33) into Eq.(54), and taking $8\pi G=1$ for
convenience, one can obtain
\begin{equation}\ddot{\delta}+2(\frac{2}{3}-\frac{\alpha}{3\beta})\frac{\dot{\delta}}{t}-\frac{1}{2}\rho_e(\frac{t}{t_e})^{-2+\frac{2\alpha}{\beta}}\delta=0\end{equation}
In the standard model, we know the solution of perturbation
$\delta\propto\widetilde{t}^{2/3}$. The relation between physical
cosmological time and conformal time is
 \begin{equation}\int{d\widetilde{t}}=\int{e^{-\frac{1}{2}\alpha\sigma}dt}\end{equation}
namely
\begin{equation}\widetilde{t}\propto t^{\frac{\beta-\alpha}{\beta}}\end{equation}
So \begin{equation}\delta\propto
t^{\frac{2}{3}-\frac{2\alpha}{3\beta}}\end{equation}
 \par Since the latter is tightly constrained from CMBR
measurements, models with slow growth of perturbation can be ruled
out. However, our model of dilaton cosmology is viable because the
growth of density perturbation in figs.1-2 is quicker than the one
in standard model. The dilaton couples weakly to ordinary
matter($\alpha^2<10^{-3}$, which seems to argue against the
existence of long-range scalars. Perhaps such a pessimistic
interpretation of the limit is premature.[5]), so the difference
between our model and $\Phi$CDM is small. Just like General
Relativity as a cosmological attractor of induced gravity[5], the
$\Phi$CDM is a cosmological attractor of our model. The
dilaton-field perturbation didn't grow during the matter-dominated
epoch. In dilaton-field dominated epoch, the presence of a
substantial amount of dilaton-field energy density inhibits the
growth of perturbations in the baryonic fluid. This is one of the
reasons why the dilaton field could have come to dominate the
Universe only recently if galaxies were formed by gravitational
instability.
\\ \vskip 0.5in
 {\noindent\Large \bf References}
\small{
\begin{description}
\item {[1]} {J.S.Bagla,H.K.Jassal and
T.Padmamabhan,Phys.Rev.D67,063504(2003);\\
           and see References 1-8 in this paper.}
\item {[2]}{L.Kofman and A.Linde, hep-th/0205121.}
\item {[3]}{D.La,Phys.Rev.D44, 1680(1991);\\
            D.La and P.J.Steinhardt, Phys.Rev.Lett 62,376(1989);\\
            D.La,P.J.Steinhardt and E.Bertschinger, Phys.Lett.B 231,232(1989).}
\item {[4]}{D.I.Kaiser,Phys.Rev.D52,4295(1995);\\
            J.Garcia-Bellido and D.wands,Phys.Rev.D52,5437(1996)}
\item {[5]}{Y.M.Cho,Phys.Rev.Lett 68,3133(1992);\\
            T.Damour and K.Nordtvedt,Phys.Rev.D48,3436(1993);\\
            T.Damour and K.Nordtvedt,Phys.Rev.Lett 70,2217(1993).}
\item {[6]}{E.J.Copeland,A.R.Liddle and D.Wands, Phys.Rev.D57,4686(1998)}
\item {[7]}{S.Weinberg,Gravitation and cosmology,John Wiley,(1972)}
\item {[8]}{B.Ratra and P.J.E.Peebles, Phys.Rev.D37, 3406(1988)}
\end{description}}
\end{document}